# Latent User Linking for Collaborative Cross-Domain Recommendation

Sapumal Ahangama and Danny Chiang-Choon Poo

School of Computing, National University of Singapore

sapumal@comp.nus.edu.sg, dannypoo@nus.edu.sg

## ABSTRACT

With the widespread adoption of information systems, recommender systems are widely used for better user experience. Collaborative filtering is a popular approach in implementing recommender systems. Yet, collaborative filtering methods are highly dependent on user feedback, which is often highly sparse and hard to obtain. However, such issues could be alleviated if knowledge from a much denser and a related secondary domain could be used to enhance the recommendation accuracy in the sparse target domain. In this publication, we propose a deep learning method for cross-domain recommender systems through the linking of cross-domain user latent representations as a form of knowledge transfer across domains. We assume that cross-domain similarities of user tastes and behaviors are clearly observable in the low dimensional user latent representations. These user similarities are used to link the domains. As a result, we propose a Variational Autoencoder based network model for cross-domain linking with added contextualization to handle sparse data and for better transfer of cross-domain knowledge. We further extend the model to be more suitable in cold start scenarios and to utilize auxiliary user information for additional gains in recommendation accuracy. The effectiveness of the proposed model was empirically evaluated using multiple datasets. The experiments proved that the proposed model outperforms the state of the art techniques.

## 1 INTRODUCTION

With the widespread adoption of information systems, recommender systems are widely used in many information systems such as e-commerce sites, social media networks and online news portals to alleviate the problem of information overload to the users and help users to find the most suited products based on their personal taste. Collaborative filtering (CF) methods are widely used in these applications to predict the most suited recommendations to users according to their similarity to other users. However, CF methods are highly dependent on user feedback data which the users are not always willing to provide due to various reasons [11]. In most scenarios, user feedback data is highly sparse and incomplete [9]. As a result, many CF methods suffer from issues of data sparsity and cold start [2].

Such issues could be alleviated if knowledge from much more denser and related secondary domains could be used to enhance the recommendation accuracy in the intended domain [2]. For example, the knowledge of user's preference in movies (source domain) could be transferred or adapted to books (target domain) to improve the recommendation quality of books. The problem is referred to as cross-domain recommendation and has received increased interest within the research community in recent times [2; 9; 10; 16; 18]. In short, cross-domain recommendation aims at adapting or transferring knowledge from data models that are trained using a source domain to be effectively used in a target domain, where source and target domains are different but related [9; 11; 16].

Prior researchers have proposed many methods for cross-domain recommendation. Matrix Factorization (MF) is considered as the most popular method until recent times [8]. MF methods project the users and items into a shared latent space represented as a vector. In the case of cross-domain scenarios, a naïve approach would be to combine the data from both the domains to apply MF on the combined dataset. A much better approach is using factorization methods that model domains jointly to bridge links between the domains [14]. However, it is noted that MF methods learn shallow features that minimize a distance metric between the domains [8; 9; 17]. Shallow methods will not be able to model complex inter domain user interaction behaviors especially in sparse environments [9]. Deep learning methods are able to identify latent factors to represent complex user interactions as well as group features hierarchically based on the relevance [17]. As a result, recent applications of deep learning methods in recommendation problems have proven to produce superior accuracies in recommendation [8; 15; 27; 28] and superior transfer capabilities for cross-domain recommendation [9; 10].

In this paper, we propose a deep learning method for cross-domain recommender systems. We take a user centric approach by linking low dimensional user latent representations across domains to create a bridge between the domains for knowledge transfer. This is due to the emergence and the ability to identify similarities of user tastes and behavioral patterns across domains with low dimensional latent representations [1]. Effective cross-domain knowledge transfer can be achieved by identifying these cross-domain similarities, which can be used as a link between the domains. We use the Variational Autoencoder (VAE) [12] as the building block of the proposed model due to its ability to generate optimal latent representations of the input data, its generative nature and the ability to identify non-linear relationships of the data [29]. However, it is required to model the domains separately to identify the domain specific distributional properties separately. Yet, it is required to model the domains jointly to identify the cross-domain links. As a result, we combine two VAEs at their embedding layer in such a manner that knowledge from source domain could be transferred to the target domain at the embedding layer. However, the model architecture allows each domain to have separate VAEs to learn the domain distributions which will be optimized jointly such that knowledge is transferred from source domain to the target domain to improve recommendation quality. Unlike the prior research which used similar autoencoder models [23; 30], we do not make the weak assumption of source and target domains being equally balanced [26]. In our model, we remove this assumption and prove that such a design is not suitable for cross-domain knowledge transfer. In addition, we include further contextualization to handle sparse cross-domain recommendation data and for better transfer of cross-domain knowledge. Apart from the generic model, two further extensions are also developed. First, a model specific for the cold start scenario due to the uniqueness of the problem and secondly, a model that incorporates auxiliary information of the user paving the way for further extensions of the model.

## 2 RELATED WORK

Recommender systems enable to alleviate the problem of information overload and help users to find the most suited products based on their personal taste. Recommender systems may be collaborative, content based [4; 16] or a hybrid of the two concepts [10; 15; 27; 28]. Unlike content based methods, collaborative filtering operates based on the user-item interactions which may be beneficial in scenarios where content is harder to obtain [9].

Prior researchers have proposed many methods for cross-domain collaborative recommendation. A recent survey has categorised the domain definitions for cross-domain recommendation at item level, item-attribute level, item-type level and at system level [2]. Until recent times, Matrix Factorization (MF) is considered as the most popular method for recommender systems in general [8]. MF methods project the users and items into a shared latent space represented as a vector. The vector like representation of users and items will be used for estimating a user's interaction with an item. In the case of cross-domain scenarios, a naïve approach would be to combine the data from both the domains to apply MF on the combined dataset [18]. Most of the cross-domain methods proposed are extended from single domain MF models with knowledge from source domains transferred into target matrix using shared factors [11]. For example, factorization that model the source and target domains jointly to bridge links between the domains are proposed [14]. Collective matrix factorization (CMF) is a method that links target domain matrix to source domain matrix to share the user factor matrix across all domains [24]. Cross-domain Triadic Factorization (CDTF) is another similar approach that models the triadic relation of user-item-domain to capture the interactions between domain-specific user factors and item factors [11]. Cross-domain Collaborative Filtering (CDCF) factorization machine based model incorporates auxiliary information from secondary domains [18]. Further methods propose to leverage rating patterns from multiple incomplete source domains to improve the quality of recommender systems [6]. However, it is noted that MF methods learn shallow features that minimize a distance metric between the domains [8; 9]. Shallow methods will not be able to model complex inter domain user interaction behaviors especially in sparse environments [9].

Deep learning methods are able to identify latent factors to represent complex user interactions as well as group features hierarchically based on the relevance [8]. As a result, recent applications of deep learning methods in recommender systems have proven to produce superior accuracies in recommendation [8; 15; 27; 28] and superior transfer capabilities for cross-domain recommendation [9; 10]. A recent study using multi-layer perceptron (MLP) proved the superior capabilities of deep learning in single domain CF [8]. Research has also been proposed where MF and MLPs are used in cross domain recommendations [19]. In this model, source domain specific user and item latent representations generated through MF are mapped to a target domain estimate using a deep neural network. Further deep learning models based on the autoencoders have shown improved results where content and other item data was combined to single domain CF to propose hybrid methods [15; 27; 28]. In the case of cross-domain recommendation, CONET has extended the MLP model to multiple domains [9]. The model was further extended to incorporate content information in a cross-domain setting to propose a hybrid model [10]. The general trend in deep learning based CF methods has been to move towards hybrid approaches by incorporating text content to improve the performance [10; 14; 17; 23; 28]. In our research, we will base our model on user-item relationship with a user centered model to show that further improvement can be achieved by proper derivation of the research model. In addition, there might be scenarios where hybrid approaches are not feasible due to non-availability of content information [9]. Yet, we also show how the proposed model could be extended to incorporate auxiliary user information too.

## 3 PROBLEM FORMULATION

Prior to introducing the problem formulation and the model in Section 4, we define some of the notations in Table 1.

Latent User Linking for Cross-Domain Recommendation

In formulating the problem, we closely follow the definitions used in recent related research [9; 10]. The entry $r_{Sij}$ ($r_{Tij}$) indicates whether the $i$th user has interacted with the $j$th item in source (target) domain with 0 indicating no interaction. The goal of cross-domain recommendation is to rank a set of items in the target domain based on the relevance of the items to a user using the information of user's target domain and source domain interaction histories. Hence, for $i$ th user the probability of interacting with an item $j$ in the target domain is defined in the proposed model as follows.

$$\hat{r}_{Tij} \triangleq p(r_{Tij} \mid r_{Ti}, r_{Si}) \quad (1)$$

The above equation indicates that the conditional probability of a user $i$ interacting with an item $j$ in the target domain is dependent on two factors. Firstly, it is based on the individual preferences of the user in the target domain and secondly it is based on the individual preferences in the source domain. As a result, the proposed neural network model can be represented as follows with $f$ being the network function.

$$\hat{r}_{Tij} = f(r_{Ti}, r_{Si} \mid \theta) \quad (2)$$

## 4 PROPOSED MODEL

### 4.1 Base Framework

Use of shared latent features to link and transfer knowledge across source and target domains is one of the most popular CF techniques [2]. The assumption is that by identifying the latent features, highly sparse data could be represented in much denser low dimensional and low loss representations, which will enable to identify high-level user preferences. Since the source and target domains are different but related, overlapping latent features representing the user preferences across domains could be identified and linked to bridge the domains.

For successful linking of the domains, the low dimensional latent representation should be optimum with the ability to encode the input data with least loss of information and it has been further noted that due to heterogeneity of items and users across-domains, it is also necessary for the model to account for domain specific distributional properties separately [11]. However, in order to link the domains effectively, the model should optimize the domains jointly for optimum knowledge transfer. Finally, the model should be able to handle sparse data, which is common to recommender systems. In the following sections, we will elaborate how each of these criteria was met for optimum cross-domain linking.

Variational Autoencoder (VAE) is a recently proposed unsupervised neural network model for data embedding [12]. VAE is an extension of the basic autoencoder. The autoencoder consists of two connected neural networks, the encoder and the decoder networks. The encoding network converts the input data to the latent dimensions which is reconstructed back to match the input using the decoder network. The loss function aims on minimizing the reconstruction error. The intention is to achieve an effective dimensional reduction of the input data, where the low dimensional representation can be used as an input to an external machine learning algorithm. VAE is a recent improvement of the autoencoder that has produced much more superior low dimensional representations [29]. Unlike the autoencoder, VAEs are generative and learn the parameters of probability distributions that model the input data with additional constraints. The learnt probability distributions could be sampled to generate the latent embedding, which is decoded by the decoder to regenerate the input data as closely as possible.

As a result, we used VAEs as the base framework to generate optimal latent representations. The input to the VAEs will be the item interaction matrix $R_S$ or $R_T$. The VAE provides an added benefit in the input reconstruction by the decoder with respect to missing data. Since the model tries to reconstruct the data as closely as possible to original input, highly relevant missing data in the target domain of the user is expected to be given a higher probability of association based on the collaborative user-item relationships learnt from the other users.

During the encoding phase of VAE, fully connected layers are utilized with nonlinear mapping function $\pi$ as follows (Eq. 3) for a total of $K$ layers for both source and target domains (for simplicity, the notations for source and target are omitted). The last layer of the encoder generates the latent representation based on $f$ that does the re-parametrization with $\varepsilon_i$ generated from a Gaussian distribution (Eq. 4).

$$h_i^1 = \pi(W^1 r_i + b^1)$$
$$h_i^k = \pi(W^k h_i^{k-1} + b^k), \quad k = 2, \ldots, K \quad (3)$$

$$z_i = f(\mu_i, \sigma_i, \varepsilon_i) \quad (4)$$

The decoder for VAE will be a mirror of the encoder with $z_i$ being the input of the first layer. The overall loss function is as follows with $H$ representing the entropy based reconstruction error ($\mathcal{L}_{recon}$), $KL$ for KL Divergence and regularization term to avoid overfitting. The heavy use of regularization has been noted in prior recommendation research and it plays a critical role to prevent overfitting [13].

$$\mathcal{L}_{vae} = \mathcal{L}_{recon} + KL(q(z_i|r_i) \parallel p(z_i)) + \mathcal{L}_{reg} \quad (5)$$
$$\mathcal{L}_{recon} = H(r_i, \hat{r}_i)$$

**Table 1: Terms and Notations**

| Symbol | Notation |
|---|---|
| $m$ | number of users ($= \|U\|$) |
| $*$ | $* \in \{S, T\}$ represents source and target domain |
| $n_*$ | number of items in source and target domains |
| $R_*$ | user-item interaction matrices |
| $i, j$ | indexing for users and items |
| $R_* = \{r_{*i}\}_{i=1}^m$ | input data |
| $\hat{R}_* = \{\hat{r}_{*i}\}_{i=1}^m$ | reconstructed data |
| $r_{*ij}$ | interaction of $i$th user with $j$th item, $\in \{0, 1\}$ |
| $h_*^k, \hat{h}_*^k$ | $k^{th}$ layer of the encoder and decoder |
| $W_*^k, \widehat{W}_*^k$ | weights of $k^{th}$ layer |
| $b_*^k, \hat{b}_*^k$ | biases of $k^{th}$ layer |
| $\mu_*, \sigma_*$ | mean and variance vectors |
| $z_*$ | latent representation |
| $L$ | latent layer dimension |
| $\theta$ | overall model parameters |

$$\mathcal{L}_{reg} = \sum_{k=1}^{K}(\|W^k\|_F^2 + \|\widehat{W}^k\|_F^2 + \|b^k\|_2^2 + \|\hat{b}^k\|_2^2)$$

## 4.2 Extension for Cross-domain Scenarios

As mentioned previously, for effective cross-domain knowledge transfer, it is required to model the domain specific distributional properties separately [11]. As a result, the proposed model incorporates two VAEs with modelling for each domain as shown in Figure 1. The two VAEs are linked at their latent layer. The need for cross-domain recommender systems arise with the target domain data being incomplete with missing data most of the times. Because of the incomplete target domain data, the target domain VAE will not have sufficient ability to model and identify the optimal latent representations. A non-optimal target domain latent representation will significantly degrade the ability of the target domain decoder to reconstruct the target domain input data. Our intuition is that, in such scenarios the related source domain data will be able to contribute and improve the target domain data modelling and reconstruction due to the shared cross-domain similarities. In order to achieve this, the target domain decoder is linked to the latent layer of the source domain to facilitate knowledge transfer from the source domain in addition to the latent layer of the target domain. That is, both source and target domain representations will contribute to the reconstruction process of the target domain decoder. However, no knowledge is transferred from the target domain to the source domain as our goal is knowledge transfer to the target domain. This results in an asymmetric cross-domain transfer model. The decoder of the target domain will be modified as in Eq. 6. However, the encoder of both the domains and the decoder of the source domain will remain as in Eq. (3).

$$z'_i = Merge(z_{Si}, z_{Ti}) \quad (6)$$

$$\hat{h}_i^1 = \pi(\widehat{W}^1 z'_i + \hat{b}^1)$$

$$\hat{h}_i^k = \pi(\widehat{W}^k \hat{h}_i^{k-1} + \hat{b}^k), \quad k = 2, \ldots, K$$

The merge operation simply concatenates the two latent vectors. The loss function for the joint model is updated as follows.

$$\mathcal{L}_{final} = \sum_{*\in\{S,T\}} \mathcal{L}_{*,vae} \quad (7)$$

Similar autoencoder designs have been used in prior cross-domain knowledge transfer studies such as linking vision and text domains [23; 30]. However, these models assume that the source and target domains are balanced and equally distributed leading to a medium solution between the source and target domains resulting in sub-optimal solutions [26]. Since the source and target domains differ in information completeness, we assume that the domains are asymmetric and the source domain is only utilized to improve the target domain while the reverse is not done as it may lead to noisier reconstruction of the source domain with incomplete target domain data. As a result, our model is one of the first attempts of applying a shared autoencoder architecture to regenerate target domain data using asymmetric transfer from the source domain in recommender systems. The model will try to identify non-linear patterns of the user-item interaction matrix in the target domain as closely as possible. As a result, while reconstructing the target domain input, any other relevant missing target domain data of the user is expected to be associated with a higher probability based on the collaborative user-item relationships learnt from the other users.

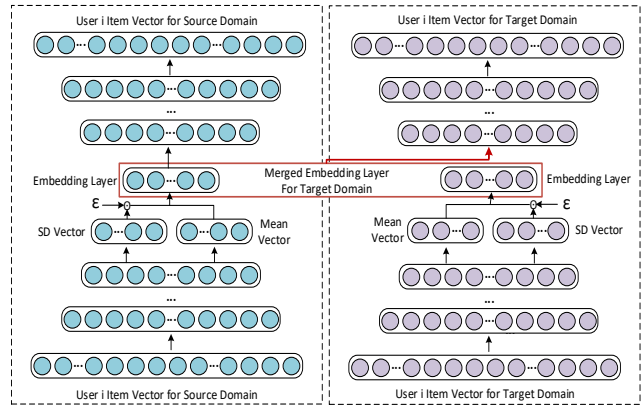

**Figure 1: Proposed Research Model for Cross-domain Data**

## 4.3 Loss Function Modifications

It is noted by prior researchers that VAEs will not handle sparse data reconstruction accurately [25]. The number of zero elements in the user-item interaction matrices far exceeds nonzero elements. In order to reduce the reconstruction loss, the model tends to give preference to zero elements over the rarer nonzero elements. Since, this is not the intended solution, it is required to impose an additional penalty for the reconstruction error of the nonzero elements [25]. The reconstruction error $\mathcal{L}_{recon}$ given in Eq. 5 is modified as in Eq. 8. $\odot$ indicates the Hadamard product with $\beta \geq 0$, a parameter to control the penalty. Since $\hat{r}_i \odot r_i$ will map all reconstructed values of zero elements to zero, $\beta H(r_i, \hat{r}_i \odot r_i)$ will purely contribute to the reconstruction of nonzero elements.

$$\mathcal{L}'_{recon} = H(r_i, \hat{r}_i) + \beta\, H(r_i, \hat{r}_i \odot r_i) \quad (8)$$

$$\mathcal{L}'_{vae} = \mathcal{L}'_{recon} + KL(q(z_i|r_i) \parallel p(z_i)) + \mathcal{L}_{reg}$$

We do intend to further modify the optimization criterion in our model to further improve the knowledge transfer from the source domain. In unsupervised knowledge transfer settings, it has shown that the similarity of marginal distributions of the high level representations facilitate knowledge transfer across domains [26]. As a result, further constraining of the model to account for domain distributional differences during training would assist in improved knowledge transfer and the model will be robust to noises as noted by prior researchers. Following the notion, we also constrained marginal distributions of the latent representations to facilitate improved knowledge transfer using Maximum Mean Discrepancy (MMD).

**Table 2. Descriptive Statistics of the Datasets**

| Dataset | #users | Source Domain | | | Target Domain | | |
|---|---|---|---|---|---|---|---|
| | | #items | #interactions | sparsity | #items | #interactions | sparsity |
| Amazon | 35,025 | 28,656 | 560,283 | 99.94% | 59,880 | 601,928 | 99.97% |
| MovieLens | 1,348 | 300 | 52,158 | 52.26% | 2,262 | 150,615 | 95.06% |

$$\mathcal{L}_{mmd} = \left\| \frac{1}{m}\sum_{i=1}^{m} z_{Si} - \frac{1}{m}\sum_{i=1}^{m} z_{Ti} \right\|_2^2 \quad (9)$$

The final loss function is updated as follows.

$$\mathcal{L}_{final} = \sum_{*\in\{S,T\}} \mathcal{L}'_{*,vae} + \mathcal{L}_{mmd} \quad (10)$$

We adopted back-propagation to minimize Eq. 10 to fine-tune the parameters. The calculation of the first term is similar to a basic VAE. The second term is derivable in a straightforward manner as adopted in prior research [26].

## 5  EVALUATION AND RESULTS

### 5.1  Experimental Setup

*5.1.1. Datasets.* Two real-world public cross-domain datasets were used for the evaluations. The descriptive statistics are given in Table 2. The first dataset is an Amazon review dataset which includes multi-domain reviews [7]. The dataset has been widely used to evaluate the performance of related cross-domain collaborative filtering research [9; 10]. The dataset corresponds to 'item level' cross-domain recommendation [2]. In our experiments, we used two popular categories, with Amazon Movies as the source domain and Amazon Books as the target domain. MovieLens 1M dataset was used as the second dataset [5]. The MovieLens dataset is much smaller and less sparser compared to the Amazon dataset. The dataset is also widely used to evaluate cross-domain collaborative filtering research [8; 22]. The dataset corresponds to 'item-attribute level' cross-domain recommendation [2]. In our experiments, we used two different combinations of movie genres as the source and target domains. Movies with genre Action were selected as the source domain whereas the combinations of Comedy, Drama, Fantasy and Romance genres were selected as the target domain. Any movie that is categorized in both the domains was disregarded in the analysis.

The ratings of data ranges from 1 to 5 stars for both the datasets with 5 stars indicating that the user shows highest positive preference on the item. Similar to prior research, we included only the ratings of 4-5 as positive interaction samples [9; 10]. The dataset was filtered to include users that have shared reviews in both the domains only.

*5.1.2. Evaluation metrics.* The leave-one-out evaluation method was adopted as the evaluation method which has been widely used in recommendation research [8-10; 21]. For each user, one interaction was held-out as the test set with the remaining data used for training. Further, we followed the common strategy which randomly samples 99 items that are not interacted by the user and ranking the held-out test item against the negative items [8-10]. Hit Ratio (HR) and Normalized Discounted Cumulative Gain (NDCG) were used as the evaluation metrics defined as follows [8; 10].

$$HR = \frac{1}{|U|}\sum_{u\in U} \delta(p_u \leq topK) \quad (11)$$

$$NDCG = \frac{1}{|U|}\sum_{u\in U} \frac{log2}{\log(p_u + 1)} \quad (12)$$

$p_u$ is the ranking for user $u$ with $\delta$ being the indicator function. HR measures whether the test item is present on the top $K$ list, with $K$ values set at $(5, 10, 20, 50)$. NDCG accounts for the position of the ranking list by assigning higher scores to hits at top ranks. In case of NDCG, the ranked list was truncated at respective $K$ values. The average value of HR and NDCG calculated across all users are reported. A higher value with lower cutoff indicates better performance.

*5.1.3. Baselines.* We compared our proposed method with the following baselines.

- BPR [21]: Bayesian personalized ranking is a popular latent factor model based on MF. The method uses ranking of pair wise loss between the positive and negative samples.
- CMF [24]: Collective matrix factorization is a multi-task learning approach which simultaneously factorizes multiple matrices of multiple domains or information sources used for cross-domain recommendation.
- MLP [8]: The model is a recent state of the art deep neural network model for collaborative filtering. The model has multiple pathways to model user and item interaction at the latent feature level.
- EMCDR [19]: The cross domain recommendation approach that derives a cross domain mapping functions for user and item features using neural nets.
- CONET [9]: Collaborative cross networks is the most recent state of the art cross-domain recommendation method. It is a deep learning model with a feed forward neural network model that models the interactions between users and multiple domains.

Since the proposed method is purely based on collaborative filtering to model user-item interactions, we limited the baselines that use the user item interaction alone similar to prior research [8; 9]. As a result, we did not include hybrid methods that incorporate other contextual or content associated with items [4; 10; 15; 16; 27; 28].

*5.1.4. Implementation.* The proposed model was implemented with Keras using TensorFlow backend[1]. The number of layers varies with different datasets. In the case of sparse Amazon dataset, the dimensions of the encoder (decoder) are $n_*$-512-256-128 (128-256-512-$n_*$) for each of the domains. For the MoveLens dataset, the dimensions of the encoder (decoder) are $n_*$-256-128 (128-256-$n_*$). Tanh function was used as the activation function for all the layers except for the last hidden layer. Since the range of values in the user-item interaction matrix is between [0, 1], we used the sigmoid function as the output function of the last hidden layer. The optimizer was Adam optimizer with initial learning rate of 0.001, with mini batch size of 128 (32) and $\beta = 40 (= 15)$ for Amazon (MovieLens) dataset. Each experiment was conducted for 100 iterations. For baseline methods, we used the code published by their respective authors and set the parameters as recommended by the authors or the best values for each dataset by fine tuning. For BPR we set learning rate at 0.05. For CMF, the weights for rating matrix were set at 75 with auxiliary matrix of 1. In case of MLP and CONET, the hidden layers for the base network were set at [64, 32, 16, 8] with batch size of 128. For EMCDR, feature space dimension was set at 20, BPR was used as the latent factor model and MLP for cross domain mapping function. For methods that are not specifically cross-domain, we merged the cross-domain data to treat the data as a single domain.

---
[1] The code is available at https://github.com/SharedVAE/sharedVAE

## 5.2 Results

Table 3 and Table 4 show a comparison of the results for the proposed method and the other baselines. As evident from the tables, the proposed method has comfortably outperformed for both the datasets in terms of HR and NDCG. The improvement is much significant in the case of the sparse Amazon dataset with an approximate increase of nearly 20% for HR at $K = 10$ compared to the next best approach, CONET. As a result, we could safely claim that proposed model has been able to utilize cross-domain knowledge to improve the results. Further experiment results are presented in Section 5.3 to validate this claim.

Several observations of further interest could be made from the results in Table 3 and Table 4. Compared among the baselines, CONET has the highest HR and NDCG for Amazon dataset. The marginal improvement of CONET over MLP is in the same range as reported in their publication. The improvement of CONET over MLP indicates the contribution of modifying the MLP model to cross-domain scenarios. The results for BPR are similar to the results for MLP. EMCDR hasn't shown a significant improvement although the model is proposed for cross domain setting. EMCDR depends of MF for latent representation of users and items. The lower results could be attributed MF being unable to model non-linear user interactions properly [8; 9; 17]. Finally, CMF is lagging behind the other baselines by a considerable margin. This can be noted in the results published by CONET for the Amazon dataset as well. CMF models source domain data as auxiliary information which is given less preference over the target domain. As a result, a wide variety of parameters were tested for CMF to improve the results. In the current context, the source domain plays an important role in contributing for recommendation accuracy. Yet by design, CMF relies more on the target domain data. Hence, the poor performance of CMF can be attributed to the over dependence on information from the target domain.

**Table 3. Evaluation Results for the Amazon Dataset**

| Method | topK = 5 | | topK = 10 | | topK = 20 | | topK = 50 | |
|---|---|---|---|---|---|---|---|---|
| | HR | NDCG | HR | NDCG | HR | NDCG | HR | NDCG |
| BPR | 0.2926 | 0.1924 | 0.4162 | 0.2392 | 0.5806 | 0.2835 | 0.8441 | 0.3381 |
| CMF | 0.1599 | 0.0917 | 0.2390 | 0.1231 | 0.3250 | 0.1467 | 0.4869 | 0.1467 |
| MLP | 0.3092 | 0.2200 | 0.4102 | 0.2484 | 0.5668 | 0.2911 | 0.7916 | 0.3329 |
| EMCDR | 0.1968 | 0.1195 | 0.3062 | 0.1597 | 0.4702 | 0.2033 | 0.7758 | 0.2655 |
| CONET | 0.3287 | 0.2298 | 0.4451 | 0.2639 | 0.6099 | 0.3169 | 0.8349 | 0.3525 |
| **Proposed** | **0.4043** | **0.2978** | **0.5336** | **0.3395** | **0.6783** | **0.3761** | **0.8832** | **0.4169** |

**Table 4. Evaluation Results for the MovieLens Dataset**

| Method | topK = 5 | | topK = 10 | | topK = 20 | | topK = 50 | |
|---|---|---|---|---|---|---|---|---|
| | HR | NDCG | HR | NDCG | HR | NDCG | HR | NDCG |
| BPR | 0.4718 | 0.3267 | 0.6105 | 0.3819 | 0.7440 | 0.4213 | 0.8998 | 0.4539 |
| CMF | 0.1884 | 0.1306 | 0.2678 | 0.1559 | 0.3820 | 0.1846 | 0.5415 | 0.2161 |
| MLP | 0.5979 | 0.4290 | 0.7611 | 0.4758 | 0.8828 | 0.5076 | 0.9829 | 0.5133 |
| EMCDR | 0.3257 | 0.2191 | 0.4562 | 0.2678 | 0.6053 | 0.3088 | 0.7982 | 0.3497 |
| CONET | 0.5845 | 0.4230 | 0.7359 | 0.4691 | 0.8716 | 0.5093 | 0.9844 | 0.5275 |
| **Proposed** | **0.6276** | **0.4553** | **0.7930** | **0.5084** | **0.9006** | **0.5358** | **0.9869** | **0.5525** |

The improvement for proposed method over CONET and MLP for the MovieLens dataset is significant yet relatively lower than for the Amazon dataset. The lower relative improvement can be attributed to higher data density and the small size of MovieLens dataset. Higher density may be an indication of data completeness. As a result, single domain methods have the ability to model the dataset with a considerable accuracy. An interesting observation that confirms this is that MLP model outperforms CONET in this setting. However, still the proposed method performs better than all the baselines. In addition, the results of EMCDR and CMF were poor compared to other baselines which can be argued as due to the same reasons mentioned previously.

## 5.3 Parameter sensitivity

In this section, we explore the effects of $\beta$ on the performance of the proposed model. $\beta$ controls the penalty of reconstructing non-zero elements incorrectly. With $\beta = 0$, the model equally reconstructs both non-zero and zero elements whereas $\beta > 0$ gives higher preference to the reconstruction of non-zero elements. Figure 2 presents the impact of HR and NDCG with varying $\beta$. As evident from Amazon dataset, at $\beta = 0$ the HR and NDCG values are significantly lower compared with $\beta > 0$. HR and NDCG values tend to be stable with the increase of $\beta$. However, the improvement of HR for MovieLens dataset in minor with $\beta > 0$. This can be attributed to the sparsity of the dataset. The Amazon dataset is highly sparse and as a result higher $\beta$ values will give preference to the reconstruction

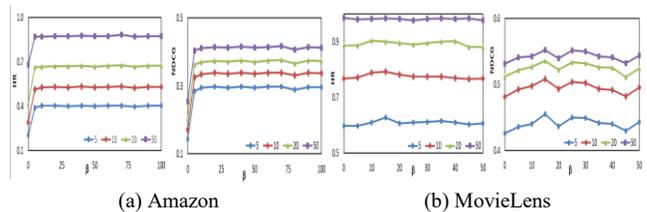

(a) Amazon  (b) MovieLens
**Figure 2: Impact of parameter $\beta$ on performance**

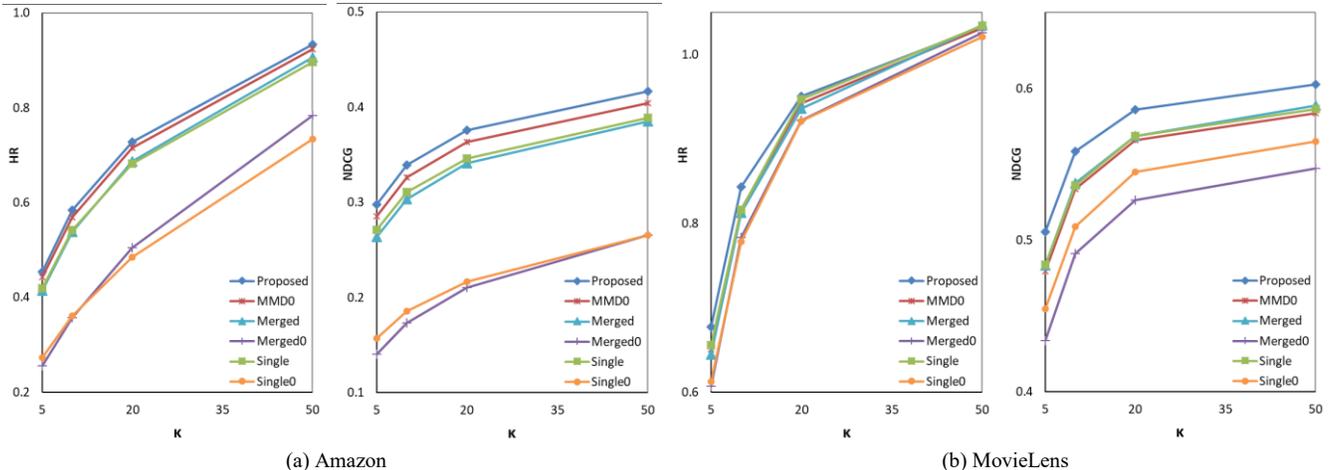

(a) Amazon  (b) MovieLens
**Figure 3: Comparison of Data Models on Recommendation Performance**

of rarer non-zero elements. Since the MovieLens dataset is less sparse, the marginal contribution of $\beta$ is less compared to Amazon dataset. Further, with the increase of $\beta$, the HR and NDCG tends to decrease for the MovieLens dataset. This can be attributed to the model tending to ignore the reconstruction of zero elements with the increase of $\beta$. In conclusion, this experiment suggests that we should control the reconstruction of non-zero elements in the model especially for sparser datasets. This observation also validates our design decision to penalize the reconstruction error of the non-zero elements.

In order to further verify our claims of knowledge transfer, several additional experiments were conducted. In these experiments, we evaluated the impact of utilizing VAEs in single domain and merged domain settings. The single domain setting will only include the data from target domain with the VAE having the structure of $n_T$-512-256-128 (128-256-512-$n_T$) for encoder (decoder) in case of Amazon dataset and the structure of $n_T$-256-128 (128-256-$n_T$) for encoder (decoder) in case of MovieLens dataset. Two variations are tested with $\beta = 0$ (referred as *Single0*) and $\beta > 0$ (*Single*). The intention of the model was to isolate the effect of data from target domain. Secondly, further two models are tested where the data was merged for both source and target domains. The VAEs will also be merged with structure of $(n_S + n_T)$-1024-512-256 (256-512-1024-$(n_S + n_T)$) for encoder (decoder) in case of Amazon dataset and the structure of $(n_S + n_T)$-512-256 (256-512-$(n_S + n_T)$) for encoder (decoder) in case of MovieLens dataset. Two variations were tested with $\beta = 0$ (*Merged0*) and $\beta > 0$ (*Merged*). The intention of the model was to isolate the effects of modelling the data distributions of the two domains separately. A further final model was also tested with the $\mathcal{L}_{mmd}$ omitted (*MMD0*) with $\beta > 0$ to isolate the effects of using MMD to constrain the marginal distributions of the two domains. For all experiments with $\beta > 0$, parameter $\beta = 40$ ($= 15$) was used for Amazon (MovieLens) dataset. The results for the experiments are presented in Figure 3. A further model for single domain was tested with $n_T$-1024-512-256 (256-512-1024-$n_T$) for encoder (decoder) in case of Amazon dataset and $n_T$-512-256 (256-512-$n_T$) for encoder (decoder) in case of MovieLens dataset for fair network and latent layer size comparison. The results were in the similar range with the respective *Single0* and *Single* models. Due to space considerations and clarity, the results are not presented.

In comparing results of *Single* and *Merged* models with the proposed model, several interesting observations could be made. First, it should be noted that *Single0* and *Merged0* show lower HR and NDCG values due to same reasons of reconstruction error. In addition, *Single* and *Merged* models are reporting much lower HR and NDCG values compared to the proposed model. This observation shows that the proposed model has better capabilities of knowledge transfer than a basic VAE. An interesting observation is that the *Single* model has better performance than the *Merged* model. This indicates that when the data from different domains are simply merged, the model is unable to identify the individual domain effects. In addition, since the source domain is much denser with information, the model will be easily biased towards the source domain. The bias will influence negatively when making predictions for the target domain. As a result, we can conclude that the assumption in existing autoencoder architectures that model target and source domains as balanced is not suitable and will lead to sub-optimal solutions [26] requiring asymmetric transfer. Further it strengthens our design decision to identify the distributional properties of each domain separately [11]. Finally, it can be observed that MMD has enabled better knowledge transfer when the proposed model is compared with *MMD0*. For example, at $K = 10$, the improvement for the Amazon (MovieLens) dataset is 2.76% (3.38%).

In training the proposed model, we set the maximum number of epochs at 100. However, it was noted that the model saturates around 20 epochs for both the datasets. Further, we tested the model with varying latent layer dimensions. It was observed that the performance increased initially with the increase of the layer size with most stable and highest results obtained at size 128. This observation is intuitive as more information could be encoded with the increase of bits in the latent layer. Further increase of latent layer size initially resulted in almost similar or lower results. However, continuous increase resulted in a significantly observable performance drop attributing to the overfitting of data with increased latent layer size.

**Table 5 Impact of Target Domain Data on Predictions**

| Target Domain | topK = 5 | | topK = 10 | | topK = 20 | | topK = 50 | |
|---|---|---|---|---|---|---|---|---|
| Data % | HR | NDCG | HR | NDCG | HR | NDCG | HR | NDCG |
| 100% | 0.4043 | 0.2978 | 0.5336 | 0.3395 | 0.6783 | 0.3761 | 0.8832 | 0.4169 |
| 75% | 0.3894 | 0.2868 | 0.5160 | 0.2700 | 0.6611 | 0.3642 | 0.8680 | 0.4051 |
| 50% | 0.3529 | 0.2578 | 0.4897 | 0.3021 | 0.6403 | 0.3402 | 0.8483 | 0.3815 |
| 25% | 0.3211 | 0.2282 | 0.4471 | 0.2687 | 0.6009 | 0.3075 | 0.8482 | 0.3565 |
| 0% | 0.2234 | 0.1536 | 0.3406 | 0.1911 | 0.4806 | 0.2263 | 0.7563 | 0.2810 |

## 6 COLD START MODEL

The cross-domain transfer model given in Figure 1 intends to improve the target domain recommendation with the assistance of source domain data. In other words, the model predictions primarily depend on the target domain data with a significant incremental contribution from the source domain data. This claim is proven when results of *Single* model and the proposed model are compared in Section 5.3. However, cold start users in the target domain is another important and common scenario which requires cross-domain recommendation. In such situations, the predictions need to be done with the dependence on the source domain observations of the user, as no target domain observations are available. Further experiments were conducted to evaluate the impact of amount of target data available during predictions. In these experiments, a test set of users (10%) were created randomly. The leave-one-out evaluation method was used at the prediction stage as before. However, the amount of target domain available at the prediction stage was randomly controlled. For example, all target domain data at prediction stage (0%) is removed in the case of cold start user scenario. The results are given in Table 5. MovieLens results are omitted due to space considerations.

As observed in Table 5, the recommendation accuracy significantly decreases with the decrease of available target domain data at the prediction stage. This is due to the model depending on target domain data for the predictions. As a result, the model needs to be modified to suit scenarios of cold start. The suggested modification is to remove the dependence of target domain data in the prediction model with sole dependence on the source domain data. It is intuitive that such a model will have a negative impact on the overall model accuracy, yet would yield improved performance during cold start. The modified training and prediction models are given in Figure 4.

**Table 6 Cold Start Evaluation for the Amazon Dataset**

| Method | topK = 5 | | topK = 10 | | topK = 20 | | topK = 50 | |
|---|---|---|---|---|---|---|---|---|
| | HR | NDCG | HR | NDCG | HR | NDCG | HR | NDCG |
| BPR | 0.1538 | 0.0920 | 0.2509 | 0.1269 | 0.4035 | 0.1671 | 0.7220 | 0.2316 |
| MLP | 0.1811 | 0.1238 | 0.2796 | 0.1583 | 0.4141 | 0.1979 | 0.6682 | 0.2313 |
| EMCDR | 0.1240 | 0.0722 | 0.2143 | 0.1039 | 0.3683 | 0.1437 | 0.7034 | 0.2118 |
| **Proposed** | **0.2706** | **0.1910** | **0.4063** | **0.2347** | **0.5514** | **0.2713** | **0.8094** | **0.3224** |

**Table 7 Cold Start Evaluation for the MovieLens Dataset**

| Method | topK = 5 | | topK = 10 | | topK = 20 | | topK = 50 | |
|---|---|---|---|---|---|---|---|---|
| | HR | NDCG | HR | NDCG | HR | NDCG | HR | NDCG |
| BPR | 0.1509 | 0.0937 | 0.2380 | 0.1249 | 0.3587 | 0.1569 | 0.5708 | 0.1996 |
| MLP | 0.3269 | 0.2168 | 0.5102 | 0.2802 | 0.7237 | 0.3470 | 0.9493 | 0.3779 |
| EMCDR | 0.0522 | 0.0251 | 0.0821 | 0.0348 | 0.1940 | 0.0605 | 0.5149 | 0.1265 |
| **Proposed** | **0.3963** | **0.2644** | **0.5801** | **0.3236** | **0.7759** | **0.3731** | **0.9590** | **0.4101** |

In the cold start model, the decoder for the target domain is modified such that it depends on an intermediary layer $z'_{Ti}$ rather than on the latent layer of the target domain encoder (Eq. 13-15). This eliminates the direct dependence of target domain data for the reconstruction of the target domain in the prediction model as shown in Figure 4b. However, the intermediate layer $z'_{Ti}$ and $z_{Ti}$ in the training model are constrained as in Eq. 16 to enforce knowledge transfer from the target domain to the source domain by accounting for a mapping loss. As a result, the final loss function given in Eq. 10 will be modified with added component of $\mathcal{L}_{ml}$.

$$z'_{Ti} = \pi(W'z_{Si} + b') \qquad (13)$$

$$\hat{h}_i^1 = \pi(\widehat{W}^1 z'_{Ti} + \hat{b}^1) \qquad (14)$$

$$\hat{h}_i^k = \pi(\widehat{W}^k \hat{h}_i^{k-1} + \hat{b}^k), \ k = 2, \dots, K \qquad (15)$$

$$\mathcal{L}_{ml} = \frac{1}{L}\sum_{l=1}^{L}(z'_{Til} - z_{Til})^2 \qquad (16)$$

The experiment results are given in Table 6 and Table 7. For evaluation, the users were randomly split test-trained at 10 - 90

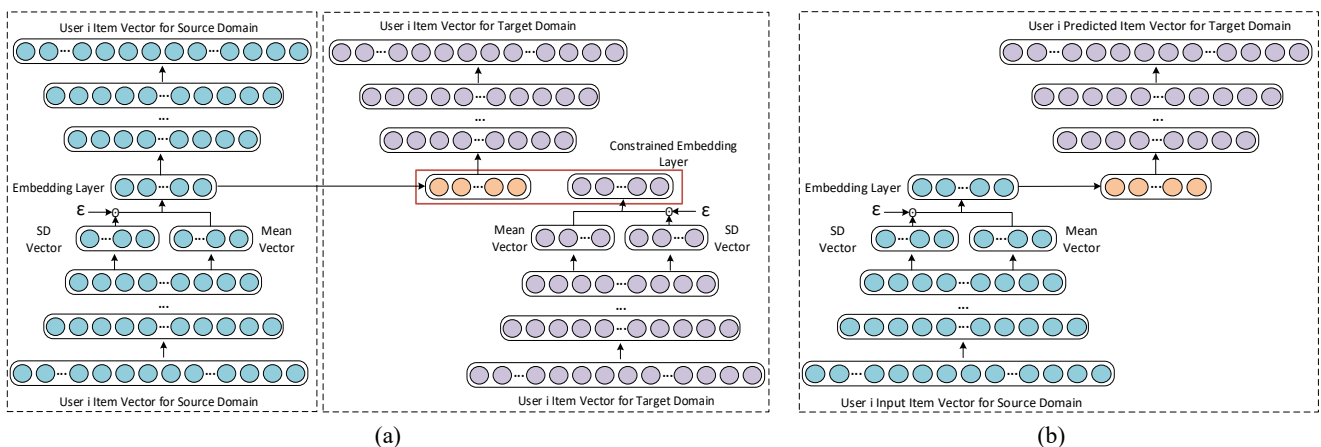

**Figure 4 Cold Start Model for Training (a) and Prediction (b)**

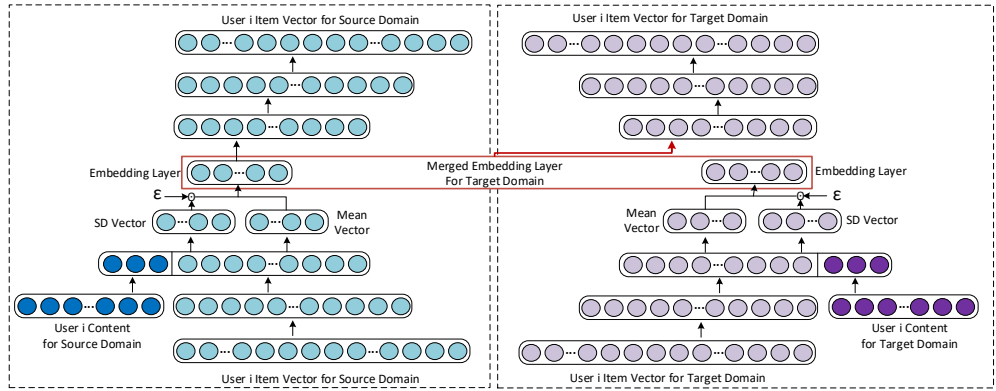
**Figure 5 Model for Incorporating User Auxiliary Information**

ratio. For test users, all the target domain observations were used as test ratings. The prediction model given in Figure 4b was used for target domain predictions. As a result, the model will not utilize any target domain observations as input at the prediction stage. The results for CONET and CMF are omitted due to issues of the methods with cold start users. CONET requires the users to be completely shared across the domains as mentioned by the authors, which is not possible with cold start users. Modifications of this requirement create very low HR and NDCG values. In case of CMF, the source domain information is considered as auxiliary information of the users. It is generally required to be a denser representation and if this requirement is not satisfied the model tends to produce very low results [3]. Due to this reason, we were not able to produce any comparable results for CMF even for a wide variety of parameter testing. As a result, the analysis for cold start users is limited to BPR, MLP and EMCDR.

As evident from Table 6 and Table 7, the proposed method has comfortably outperformed for both the datasets in terms of HR and NDCG for cold start users in the target domain. The improvement is much significant in the case of the sparse Amazon dataset with an approximate increase of nearly 36% for HR at $K = 10$ compared to the next best approach, MLP. Since the proposed model relies only on source domain data at the prediction stage, we could safely claim that proposed model has been able to learn cross-domain transfer model to improve the cold start user results. It should also be noted that the generic model given in Figure 1 still outperforms all the baselines in cold start even without the modification as per Table 5 (0% results).

The results also indicate that whenever a considerable amount of target domain data is available for the user, it is better to use the generic model to achieve better recommendation accuracy as the target domain data will contribute to a performance gain.

## 7   USER AUXILIARY INFORMATION

As mentioned in the introduction, our approach for cross-domain recommendation is centered on user modeling or user latent representation. Apart from the user-item interaction matrix, recommender systems are also available with additional auxiliary information about the users. For example, user profile descriptions or other user generated content such as review text. This auxiliary content may contain vital information that could further improve the user latent representation leading to an improved transfer model. It is important that the model could be extended to incorporate this information. We further extend the model to include these auxiliary content as shown in Figure 5.

In the model, we treat the user created review text content as user auxiliary information. The text content is first converted to a vector representation using Doc2Vec [20]. Any other auxiliary content can similarly be fed to the model provided the content is converted to a suitable vector format. As shown in Figure 5, we propose the incorporation of auxiliary information as a separate sub-encoder which is merged with the final layer of the main encoder. A separate sub-encoder is adopted due to the sparsity differences of user-item interaction matrix (usually sparse) and the vector representation of auxiliary information (usually denser). If both the data types are simply merged at the input level, the model will give higher preference to the denser auxiliary information representation creating a bias. As a result, the data is merged at a higher layer of the encoder which has already down-sampled the input. In addition, this allows the identification of data type specific properties independently. Further, the decoder will not regenerate the auxiliary information, as our intention is to regenerate the item-interaction matrices as accurately as possible with the contributions from auxiliary information.

The results for the model are given in Table 8 for Amazon dataset. MovieLens results are omitted as the dataset doesn't include any auxiliary user information. The model had same network structure and parameters as with proposed model in Figure 1 with dimensions for Doc2Vec representation were set at 256 and the sub-encoder had two layers of structure 256-128. As evident from the table, the recommendation accuracy has significant increase with the incorporation of user auxiliary information. This is validating our intuition to include user auxiliary information in the model. The cold start model can also be extended in a similar manner to incorporate auxiliary user information.

**Table 8 Impact of Auxiliary Information on Predictions**

| Auxiliary Information | $topK = 5$ | | $topK = 10$ | | $topK = 20$ | | $topK = 50$ | |
|---|---|---|---|---|---|---|---|---|
| | HR | NDCG | HR | NDCG | HR | NDCG | HR | NDCG |
| No | 0.4043 | 0.2978 | 0.5336 | 0.3395 | 0.6783 | 0.3761 | 0.8832 | 0.4169 |
| Yes | 0.4612 | 0.3431 | 0.5921 | 0.3853 | 0.7341 | 0.4212 | 0.9124 | 0.4568 |

## 8  CONCLUSION

In this paper, we have shown that our model can achieve state-of-the-art cross-domain collaboration recommendation accuracy. We followed a user centric latent modeling approach with deep knowledge transfer to achieve this. To best of our knowledge, this is the first attempt of model knowledge transfer in cross-domain recommendation with an account of asymmetry of the source and target domains in recommender systems. Further, we identified unique conditions associated with recommendation data to further enhance the optimization criterion to improve the results of the proposed method. In addition, we presented further enhancements to the model to suite cold start scenario and to scenarios available with user auxiliary information. In order to empirically validate our claims, we conducted experiments against current state of the art methods using two public benchmark datasets. Further experiments were conducted to prove the superior cross-domain knowledge transfer capabilities.

Although the model was tested for two domains, the model could be easily extended to separately model multiple source domains as well. In our experiments, both source and target domain networks followed similar network structure and parameter settings. However, it could be expected to further improve the results if the network structure and parameters were fine-tuned separately for each of the domains. We intend to work in several directions in the future to further improve the model. Much of the recent deep learning based collaborative filtering algorithms have resorted to proposing hybrid approaches by incorporating item information to improve the results [10; 15; 16; 27; 28]. Following this direction, we do intend to extend the model by incorporating the item view point in addition to the current user centered view. This can be achieved using multi view VAE architectures. In addition, the model could be further constrained to incorporate the proximity of users by considering the user overlaps to further improve the results. We leave these extensions for future research.